\documentclass[12pt]{article}
\usepackage{amsfonts,amsmath,amssymb}
\usepackage{amsthm,amscd}
\makeatletter

\@addtoreset{equation}{section}
\makeatother
\theoremstyle{definition}
\newtheorem*{pr}{Proposition}
\newtheorem*{rem}{Remark}
\newtheorem*{pproof}{Proof}

\newcommand{\bib}[2]{\frac{\partial {#1}}{\partial {#2}}}

\newcommand{\nyoro}{$\sim$}
\newcommand{\sdiff}{\mathop{\rm SDiff}\nolimits}
\newcommand{\diag}{\mathop{\rm diag}\nolimits}

\begin{document}
\title{Anti-self-dual Maxwell solutions on hyperk\"ahler manifold 
and $N=2$ supersymmetric Ashtekar gravity}
\author{
\thanks{E-mail address : ootsuka@sci.osaka-cu.ac.jp} Takayoshi OOTSUKA ,
\thanks{E-mail address : miyagi@sci.osaka-cu.ac.jp} Sayuri MIYAGI , \\
\thanks{E-mail address : yasui@sci.osaka-cu.ac.jp} Yukinori YASUI \   and
\thanks{E-mail address : zeze@sci.osaka-cu.ac.jp} Shoji ZEZE  \\
{\small Department of Physics, Osaka City University, 
Sumiyoshiku, Osaka, Japan } }
\date{}
\maketitle


\begin{abstract}

Anti-self-dual (ASD) Maxwell solutions on 4-dimensional hyperk\"ahler
manifolds are constructed. The $N=2$ supersymmetric half-flat equations are
derived in the context of the Ashtekar formulation of $N=2$ supergravity.
These equations show that the ASD Maxwell solutions have a direct connection
with the solutions of the reduced N=2 supersymmetric ASD Yang-Mills
equations with a special choice of gauge group. Two examples of the Maxwell 
solutions are presented. 

\quad \\
\quad \\
PACS numbers 04.20.Jb, 04.65.+e
\end{abstract}


\section{Introduction}

The Ashtekar formulation of Einstein gravity gives a new insight to 
the search for anti-self-dual (ASD) solutions without cosmological
constant. These are constructed
from the solutions of certain differential equations for volume-preserving
vector fields on a 4-dimensional manifold. 
This characterization of the ASD solutions has been originally given by
Ashtekar, Jacobson and Smolin~\cite{A-J-S}, and further elaborated by Mason and
Newman~\cite{M-N}. In the following, we call their differential equations the half-flat
equations. These equations clarify the relationship between
the ASD solutions of the Einstein and the Yang-Mills equations. 
Indeed, if we specialize the gauge group to be a volume-preserving diffeomorphism
group, the reduced ASD Yang-Mills equations on the Euclidean space are identical to
the half-flat equations~\cite{M-N}.

Looked at geometrically, the Ashtekar formulation emphasizes the hyperk\"ahler
structures that naturally exist on ASD Einstein solutions. A hyperk\"ahler manifold
is a 4$n$-dimensional Riemannian manifold $(M,g)$ such that $(1)$ $M$ admits three complex
structures $J^a (a=1,2,3)$ which obey the quarternionic relations
 $J^a J^b =-\delta_{ab}-\epsilon_{abc} J^c$; 
$(2)$ the metric $g$ is preserved by $J^a$;$(3)$ the $2$-forms $B^a$ defined by 
$B^a (X,Y)=g(J^a X,Y)$ for 
all vector fields $X,Y$ are three k\"ahler forms, i.e. $dB^a=0\quad (a=1,2,3)$. 
The solutions of the half-flat
equations ensure the conditions above and hence 4-dimensional 
hyperk\"ahler metrics are ASD Einstein solutions. 

Recently, making use of 
the half-flat equations we have explicitly constructed several 
hyperk\"ahler metrics~\cite{Hashi}.  
Subsequently here we extend the half-flat equations to the case of
$N=2$ supergravity \footnote{$N=1$ half-flat equations are obtained from our equations 
(\ref{eq:hf2}) \nyoro (\ref{eq:Liedel}) by putting $T=0$.} 
Our formulation has
the advantage that the setting of $N=2$ supersymmetric Yang-Mills theory is
automatically provided. In particular  ASD Maxwell solutions on hyperkh\"aler manifolds are 
elucidated through the relationship to the reduced $N=2$ ASD Yang-Mills equations.
In the literature~\cite{K-N-G,G-N-K} the $N=2$ ASD supergravity has been
investigated by using the superfield formulation, but our approach is
very different and the results in the present work are more concrete. 

In Section 2 we review the half-flat equations. In Section 3
we present a new construction of ASD Maxwell solutions on hyperk\"ahler
manifolds and derive the $N=2$ supersymmetric half-flat equations. 
Finally, in Section 4 two examples of ASD Maxwell solutions are given.

The following is a summary of the notation used in this paper.
The $so(3)$ generators and the Killing form are denoted by $E_{a}(a=1,2,3)$ and
 $\langle \, , \, \rangle$, respectively. 
The symbols  ${\eta}^{a}_{\mu \nu}$ and 
${\bar{\eta}^a}_{\mu \nu}$ $(a=1,2,3; \mu,\nu=0,1,2,3)$ 
represent the 't Hooft matrices satisfying the relations: 
\begin{equation}
{\eta}^{a}_{\mu \nu} = - \frac{1}{2}
 \epsilon_{\mu\nu\lambda\sigma} {\eta}^{a}_{\lambda \sigma},
\quad 
  {\bar{\eta}}^{a}_{\mu \nu} =  \frac{1}{2} 
\epsilon_{\mu\nu\lambda\sigma} {\bar{\eta}}^{a}_{\lambda \sigma}
\end{equation}
and
\begin{equation}\label{'tHooft}
   {\eta}^{a}_{\mu \nu}{\eta}^{b}_{\mu \sigma}
  = {\delta}_{ab} {\delta}_{\nu \sigma} + {\epsilon}_{abc}
  {\eta}^{c}_{\nu \sigma}  \quad (\bar{\eta}^a_{\mu \nu} 
\, \mbox{satisfy the same relations.}).
\end{equation}
In Section 3 we consider a space-time with the signature $(++--)$. 
Then the metrics $\hat{g}_{\mu\nu}= \diag (1,1,-1,-1)$ and
 $\kappa_{ab}=\diag (1,-1,-1)$ are used to lower 
and rise the indices of ${\eta}^{a}_{\mu \nu}$ $({{\bar{\eta}}^a}_{\mu \nu})$.

\section{Half-flat~equations}

In this section, we briefly describe the 4-dimendional hyperk\"ahler geometry 
from the point of view of Ashtekar gravity~\cite{A-J-S,M-N}.
We use the metric of the Euclidean signature for avoiding complex variables.
The Ashtekar gravity consists of an $so(3)$ connection 
1-form $A = A^a \otimes E_a$ and an $so(3)$-valued 2-form $B = B^a \otimes E_a$
on a 4-dimensional manifold $M$.
The action is given by~\cite{C-D-J-M}
 \begin{equation}\label{Ash}
  S_{Ash} = \int_{M} \langle B \wedge F \rangle 
          - \frac12 \langle C(B) \wedge B \rangle,  
 \end{equation}
where $F = dA + \frac12 [A \wedge A]$, $C(B)= {C^a}_b B^b \otimes E_a$ and 
$C={C^a}_b E_a \otimes E^b$ is a Lagrange multiplier field which obeys the conditions,
${C^a}_b={C^b}_a$ and ${C^a}_a=0$.
The equations of motion are 
\begin{eqnarray}
  &F - C(B) = 0 \label{eq:Ash1}\\
  &DB = 0 \label{eq:Ash2}\\
  &B^1 \wedge B^2 = B^2 \wedge B^3 = B^3 \wedge B^1 = 0 \label{cnsB1}\\
  &B^1 \wedge B^1 = B^2 \wedge B^2 = B^3 \wedge B^3 \label{cnsB2},
\end{eqnarray}
where $D$ is the covariant derivative with respect to $A$.
The algebraic equations (\ref{cnsB1}) and (\ref{cnsB2}) represent the constraints
of this system.

To solve the constraints we introduce linearly independent four vector fields 
$V_{\mu}\,(\mu=0,1,2,3)$ and a volume form $\omega$ on $M$.
Then the solutions become the self-dual 2-forms
\begin{equation}\label{eq:BtoV}
  B^a = \frac12 {{\bar{\eta}}^a}_{\mu \nu} \iota_{V_{\mu}} 
      \iota_{V_{\nu}} \omega ,
\end{equation}
where $\iota_{V_{\mu}}$ denotes the inner derivation with respect to $V_{\mu}$.
We proceed to solve the remaining equations (\ref{eq:Ash1}) and (\ref{eq:Ash2}).
For the hyperk\"ahler geometry, which we will forcus on in this paper,
$C$ must be taken to be zero because ${C^a}_b$ are the coefficients of self-dual Weyl curvature;
this is equivalent to the requirement that the holonomy group is contained 
in subgroup Sp(1) of SO(4).
With this choice, (\ref{eq:Ash1}) becomes $F=0$  and if we take the gauge
fixing $A=0$, (\ref{eq:Ash2}) reduces to 
\begin{equation}
  dB^a = 0 \quad (a=1,2,3).
\end{equation}
Thus (\ref{eq:BtoV}) implies the half-flat equations~\cite{A-J-S,M-N},
\begin{eqnarray}
&\frac{1}{2} {{\bar{\eta}}^a}_{\mu \nu} [V_{\mu},V_{\nu}] = 0, \label{eq:MN1} \\  
&L_{V_{\mu}}\omega = 0.  \label{eq:MN2}
\end{eqnarray}
This can be seen by applying the formula :
\begin{equation}
d(\iota_X \iota_Y \alpha)=\iota_{[X,Y]} \alpha + \iota_Y L_X \alpha -\iota_X L_Y \alpha
+ \iota_X \iota_Y d\alpha \label{ii}
\end{equation}
for vector fields $X,Y$ and a form $\alpha$.
Given a solution of (\ref{eq:MN1}) and (\ref{eq:MN2}), we have a metric
\begin{equation}
  \label{eq:hk-metric}
  g(V_{\mu},V_{\nu}) = \phi \delta_{\mu \nu} ,
\end{equation} 
where $\phi=\omega(V_0,V_1,V_2,V_3)$.
This metric is invariant by the three complex structures
\begin{equation}
  \label{eq:hk-st}
  J^{a} (V_{\mu}) = {{\bar{\eta}}^a}_{\nu \mu} V_{\nu} \quad (a=1,2,3),
\end{equation}
which obey the relations 
$J^a J^b = - \delta_{ab} - \epsilon_{abc}J^c$
and give the three K\"ahler forms $B^a(V_{\mu},V_{\nu})=g(J^a(V_{\mu}),V_{\nu})$.
Thus the triplet $(M, g, J^a)$ is a hyperk\"ahler manifold.
Conversely, it is known that every 4-dimensional hyperk\"ahler manifold 
locally arises by this construction~\cite{M-N,Don}.

This formulaton yields that the vector fields $V_{\mu}$ may be identified with
the components of a space-time independent ASD Yang-Mills connection on 
$\mathbb{R}^4$.
Indeed, (\ref{eq:MN2}) is the assertion that the gauge group is the diffeomorphism
group $\sdiff_\omega (M)$ preserving the volume form $\omega$,
and (\ref{eq:MN1}) are explicitly written as
\begin{eqnarray}
  \label{eq:MN0}
  &[V_0, V_1] + [V_2, V_3] = 0 \\
  &[V_0, V_2] + [V_3, V_1] = 0 \\
  &[V_0, V_3] + [V_1, V_2] = 0,
\end{eqnarray}
which are equivalent to the reduced ASD Yang-Mills equations~\cite{M-N}.


\section{N=2 supersymmetric Ashtekar gravity}

We start with the chiral action for $N=2$ supergravity without 
cosmological constant~\cite{Kun-San,Ezawa}. 
The bosonic part, which is the chiral action of Einstein-Maxwell theory, 
contains a $U(1)$ connection 1-form
$a$ and a 2-form $b$ in addition to $A,B$ in (\ref{Ash}) \cite{Robin}. 
The fermionic
fields (two gravitino fields) are expressed by Weyl spinor 1-forms $\psi^i$ 
and Weyl spinor 2-forms $\chi^i$, where $i (=1,2)$ is a Sp(1) index 
representing the two supersymmetric charges. 
By using the 2-component spinor notation, the chiral action is written as
\footnote{We have re-named the variables in~\cite{Ezawa} as 
$(A_{AB},A,{\psi_{\alpha}}^A,\Sigma^{AB},B,{\chi_{\alpha}}^A,
  \Psi_{ABCD},{\kappa^{\alpha}}_{ABC},\phi_{AB})$ \mbox{$\mapsto$}
$(A_{AB},a,\frac1{\sqrt{2}}{\psi_{i}}^A,
 -B^{AB},-\frac12 b,-\sqrt{2}{\chi_i}^A,
  -C_{ABCD},-\frac1{\sqrt{2}}{\kappa^{i}}_{ABC},-\frac12 H_{AB})$}.

\begin{eqnarray}\label{N=2Ash}
&&S_{Ash}^{N=2} = \int  B^{AB} \wedge F_{AB} + b \wedge f
+ {\chi^{i}}_{A} \wedge D {\psi_i}^{A} - \frac{1}{2} b \wedge b 
- \frac{1}{8} b \wedge {\psi_i}^{A} \wedge {\psi^i}_{A} \nonumber \\ 
&&- \frac{1}{2} C_{ABCD} B^{AB} \wedge B^{CD} 
-{\kappa^i}_{ABC} B^{AB} \wedge {\chi_i}^{C} 
-\frac{1}{2} H_{AB} (B^{AB} \wedge b - {\chi_i}^A \wedge {\chi^{iB}})                 
\end{eqnarray}   
where $f=da$, and $C_{ABCD},{\kappa^i}_{ABC}$ and $H_{AB}$ 
are totally symmetric Lagrange multiplier fields. 

Let us forcus on ASD solutions. Then we can put $A_{AB}=C_{ABCD}=0$ as stated
in Sect.2, and further impose the conditions 
$H_{AB}={\kappa^i}_{ABC}={\psi_i}^A=0$.
It should be noticed that these restrictions preserve the $N=2$ 
supersymmetry; as we will see in Sect.\ref{sec:N=2half} this symmetry is properly
realized in the $N=2$ supersymmetric ASD Yang-Mills equations with the
gauge group $\sdiff_{\omega}(M)$. Now the equations of motion
derived from $S_{Ash}^{N=2}$ reduce to
\begin{eqnarray}
        &f = b                  \label{eq:N=2Pleban1} \\  
        &d B^{AB}=d b =  d {\chi_i}^A = 0 \label{eq:N=2Pleban2} \\
        &B^{(AB} \wedge B^{CD )} = 0 \label{eq:N=2Pleban3} \\
        &B^{(AB} \wedge {\chi_i}^{C )} = 0 \label{eq:N=2Pleban4} \\
        &B^{AB} \wedge b - {\chi_i}^{A} \wedge \chi^{iB} = 0.\label{eq:N=2Pleban5}
\end{eqnarray}


\subsection{Maxwell solutions on hyperk\"ahler manifolds}

We first consider the bosonic sector $(b=f,B)$ in a space-time with the Euclidean signature.
The relevant equations are obtained from
(\ref{eq:N=2Pleban1})\nyoro (\ref{eq:N=2Pleban5}) by putting ${\chi_i}^A=0$. In the previous
section we have seen that the solutions $B^a (a=1,2,3)$ are self-dual
K\"ahler forms on a hyperk\"ahler manifold $M$. Thus the equations (\ref{eq:N=2Pleban2}) and 
(\ref{eq:N=2Pleban5}) imply that $b$ is an ASD closed 2-form (ASD Maxwell
solution) on $M$. The following proposition holds.

\pr
Let $M$ be a hyperk\"ahler manifold expressed by linear independent vector
 fields $V_\mu (\mu=0,1,2,3)$ and a volume form $\omega$ as mensioned in
 (\ref{eq:MN1}) and (\ref{eq:MN2}).
If the vector field $T=T_\mu V_\mu$ satisfies
\begin{eqnarray}
&   L_{T} \omega = 0,   \label{eq:liedelT} \\
&    [V_{\mu}, [V_{\mu}, T]] = 0, \label{eq:VVT} 
\end{eqnarray}
then $b$ defined by
\begin{equation}
    b = \frac12 b^a {\eta}^{a}_{\mu \nu}\iota_{V_{\mu}} \iota_{V_{\nu}} \omega 
\quad \mbox{for}\quad b^a={\eta}^{a}_{\mu \nu}V_\mu T_\nu, \label{eq:btoVT}
\end{equation}
is an ASD closed 2-form on $M$.

\pproof
The ASD condition of $b$ immediately follows from
(\ref{eq:btoVT}). Therefore it suffices to prove that $b$ is a closed
2-form. Using the identity of the 't Hooft matrices
\begin{equation}
{\eta}^{a}_{\mu\nu}{\eta}^{a}_{\lambda\sigma}=\delta_{\mu\lambda}\delta_{\nu\sigma}
-\delta_{\mu\sigma}\delta_{\nu\lambda}-\epsilon_{\mu\nu\lambda\sigma},
\label{eq:'tHooft} 
\end{equation}
we rewrite (\ref{eq:btoVT}) in the form,
\begin{equation}
b=\iota_{V_{\mu}} \iota_{V_{\nu}}L_{V_{\mu}}(T_\nu \omega)-
 \frac{1}{2}\epsilon_{\mu\nu\lambda\sigma}
\iota_{V_{\mu}} \iota_{V_{\nu}}L_{V_{\lambda}}(T_\sigma \omega).
\label{eq:smallb2} 
\end{equation}
Let us define the vector fields
\begin{equation}
W_{\mu\nu}=[V_\mu,V_\nu]+\frac{1}{2}\epsilon_{\mu\nu\lambda\sigma}
[V_\lambda,V_\sigma].\label{eq:largeW} 
\end{equation}
Then,
\begin{equation}
b+\iota_{V_\mu} \iota_{W_{\mu\nu}}T_\nu \omega=
\iota_{V_\mu}L_{V_\mu}\iota_T \omega +
\frac{1}{2}\epsilon_{\mu\nu\lambda\sigma}
\iota_{V_\mu}L_{V_\lambda}\iota_{V_\sigma}(T_\nu \omega).\label{eq:proof3}
\end{equation}
The exterior derivative of (\ref{eq:proof3}) is evaluated as follows:
Since both the vector fields $V_\mu$ and $T$ preserve the volume form
$\omega$, we have 
\begin{eqnarray}
d(\iota_{V_\mu}L_{V_\mu}\iota_T \omega)& = & L_{V_\mu}L_{V_\mu}\iota_T \omega 
\nonumber \\
& = & \iota_{[V_\mu,[V_\mu,T]]}\omega\label{eq:proof4} 
\end{eqnarray} 
and
\begin{eqnarray}
d(\epsilon_{\mu\nu\lambda\sigma}
\iota_{V_\mu}L_{V_\lambda}\iota_{V_\sigma}(T_\nu \omega)) &=& 
\frac{1}{2}\epsilon_{\mu\nu\lambda\sigma}
(L_{V_\mu}L_{V_\lambda}\iota_{V_\sigma}-\iota_{V_\mu}L_{V_\lambda}L_{V_\sigma})(T_\nu \omega)\nonumber \\  
&=& \frac{1}{4}\epsilon_{\mu\nu\lambda\sigma}\iota_{[[V_\mu,V_\lambda],V_\sigma]}
T_\nu \omega = 0.
\end{eqnarray}
We thus find
\begin{equation}
d(b+\iota_{V_\mu}\iota_{W_{\mu\nu}} 
T_\nu \omega)=\iota_{[V_\mu , [V_\mu ,T]]}\omega.
\end{equation}
Finally, making use of (\ref{eq:MN1}), i.e. $W_{\mu\nu}=0$, combined with the
condition (\ref{eq:VVT}), we obtain the required formula $db=0$. \qed

\rem
Using the hyperk\"ahler metric (\ref{eq:hk-metric}), we can rewrite (\ref{eq:btoVT}) as
\begin{equation}
b = dg (T,\,) + \iota_{[V_{\mu},[V_{\mu}, T]]} \omega.
\end{equation}
This expression is convenient to the explicit calculation in Sect.4.

\subsection{$N=2$ supersymmetric half-flat equations}\label{sec:N=2half}

Let us return to the equations
(\ref{eq:N=2Pleban1})\nyoro (\ref{eq:N=2Pleban5}) and assume a space-time 
with the signature $(++--)$. 
It is known that the hyperk\"ahler manifolds with this signature provide 
the consistent backgrounds for closed N=2 strings~\cite{O-V1,O-V2}.
We follow the paper for the spinor notation of~\cite{K-N-G};
the spinor indices $A,B,C \cdots$ in (\ref{N=2Ash}) are 
replaced by the dotted indices $\dot{A},\dot{B},\dot{C} \cdots$. 
To solve the constraints we introduce spinor valued vector
fields $V_{i}^{A}$ in addition to the vector fields $V_\mu$
(or $V_{A\dot{B}}$) and $T$. 
Referring to (\ref{eq:btoVT}), we put
\begin{eqnarray}
  &{\chi_{i\dot{A}}} = \iota_{V_{B\dot{A}}} \iota_{{V_i}^B} \omega,  \\ \label{eq:chitoV1}
  & b = \frac12 ({\eta^a}^{\lambda \sigma} V_{\lambda} T_{\sigma})
          {\eta_a}^{\mu \nu} \iota_{V_{\mu}} \iota_{V_{\nu}} \omega 
   + \iota_{{V^i}_A} \iota_{{V_i}^A}\omega, \label{eq:chitoV2}
\end{eqnarray}
together with (\ref{eq:BtoV}), i.e. $B_{\dot{A}\dot{B}}= \frac{1}{2}
\iota_{V_{C\dot{A}}} \iota_{{V^C}_{\dot{B}}} \omega$ 
in the spinor notation (See Sect.1 for the 't Hooft matrices.). 
It is easily confirmed that these formulas
automatically satisfy
(\ref{eq:N=2Pleban3})\nyoro (\ref{eq:N=2Pleban5}). Furthermore
(\ref{eq:N=2Pleban2}) requires the following equations for the vector
fields, which are proved in a similar fashion to the preceding
proposition:
\begin{eqnarray}\label{eq:hf2}
  &\frac12 {\bar{\eta}}^{a\mu\nu} [V_\mu,V_\nu]=0 \\ \label{eq:2even}
  &[ V^{\mu} , [ V_{\mu} , T ]] + [ {V^i}_A , {V_i}^{A}] = 0 \\ \label{eq:3even}
  &[ V_{B\dot{A}} , {V_i}^B] = 0\\ \label{eq:oddeven}
  & \mbox{and}\nonumber \\
  &L_{V_\mu} \omega=L_{{V_i}^A} \omega =  L_T \omega = 0.  \label{eq:Liedel}
\end{eqnarray}

This result is satisfactory in that it gives the direct correspondence
between the ASD solutions of the $N=2$ supergravity and the $N=2$
supersymmetric Yang-Mills theory; the equations
(\ref{eq:hf2})\nyoro (\ref{eq:Liedel}) can be regarded as $N=2$
supersymmetric extension of the half-flat equations. To say more
precisely, let us recall the $N=2$ ASD Yang-Mills equation in a flat
space-time with the signature $(++--)$~\cite{K-N-G,G-N-K}.
The $N=2$ Yang-Mills theory has the field content 
$(A_\mu,{\lambda_{iA}},\tilde{\lambda}_{i\dot{A}},S,\widetilde{S})$, 
where ${\lambda_{iA}}$ and $\tilde{\lambda}_{i\dot{A}}$ 
are chiral and anti-chiral
Majorana-Weyl spinors, while the fields $S$ and $\widetilde{S}$ are real scalors.
All the fields are in the adjoint representation of gauge group. 
By the supersymmetric ASD conditoin, i.e. $\widetilde{S}=0$, the equations of motion
reduce to
\begin{eqnarray}
  \label{eq:N=2ASD-YM}
  &\frac12  {\bar{\eta}}^{a\mu \nu} [ D_{\mu} , D_{\nu}] = 0 \\
  &D^{\mu} D_{\mu} S +[ {\lambda^{i}}_{A} ,{\lambda_{i}}^{A} ] = 0 \\
  &(\sigma^{\mu} D_{\mu})_{B\dot{A}} {\lambda_i}^{B} = 0,\label{eq:lambda}
\end{eqnarray}
where $D_\mu=\partial_\mu +[A_\mu,\ ]$. 
If we require thet the fields are all constant on the space-time, and
further choose the gauge group as $\sdiff_\omega (M)$, then the equations
(\ref{eq:N=2ASD-YM})\nyoro (\ref{eq:lambda}) just become the $N=2$ supersymmetric half-flat
equations (\ref{eq:hf2})\nyoro (\ref{eq:Liedel}) with the identification 
$A_\mu=V_\mu ,\, {\lambda_i}^A={V_i}^A$ and $S=T$.


\section{Examples of ASD Maxwell solutions}

As an application of the proposition, we present two examples of ASD
Maxwell solutions on 4-dimensional hyperk\"ahler manifolds with one
isometry generated by a Killing vector field
$K=\frac{\partial}{\partial \tau}$. The first example gives the
well-known Maxwell solution and the second one leads to a new solution as
far as the authors know. We use local coordinates
$(\tau,x^1,x^2,x^3)$ and a volume form $\omega=d\tau\wedge dx^1\wedge
dx^2\wedge dx^3$ for the background manifold.

\subsection{Gibbons-Hawking background}

In this case we choose the vector fields $V_\mu$ as~\cite{Joy}
\begin{eqnarray}
V_0 &=& \phi \bib{}{\tau}, \\  
V_i &=& \bib{}{x^i} + \psi_i \bib{}{\tau} \quad (i=1,2,3),
\end{eqnarray}
where the functions $\phi,\psi_i$ are all independent of $\tau$. 
Then these vector fields preserve the volume form 
$\omega$ and (\ref{eq:MN1}) implies 
\begin{equation}
* d \phi = d \psi,
\end{equation}
where $\psi=\psi_i dx^i$ and $*$ denotes the Hodge operator on
$\mathbb{R}^3=\{(x^1,x^2,x^3) \}$. The resultant metric is the
Gibbons-Hawking multi-center metric~\cite{G-H},
\begin{equation}
ds^2=\phi^{-1} (d\tau + \psi)^2 +\phi dx^i dx^i. 
\end{equation}
The Killing vector field $T=K$ clearly satisfies (\ref{eq:liedelT}) 
and (\ref{eq:VVT}).
Applying the proposition, we have an ASD Maxwell solution~\cite{E-H},
\begin{equation}
b=da \quad \mbox{with} \quad a=\phi^{-1} (d\tau-\psi).
\end{equation}


\subsection{Real heaven background}

We choose the vector fields $V_{\mu}$ as~\cite{Hashi}
\begin{eqnarray}\label{sympV}
  V_0 &=& e^{\frac{\psi}{2}}\left(\partial_3 \psi\cos \left(\frac{\tau}{2}\right) \bib{}{\tau}
+ \sin \left(\frac{\tau}{2}\right) \bib{}{x^3}\right)\\ 
  V_1 &=& e^{\frac{\psi}{2}}\left(-\partial_3 \psi \sin \left(\frac{\tau}{2}\right) \bib{}{\tau}
+ \cos \left(\frac{\tau}{2}\right) \bib{}{x^3}\right)\\
  V_2 &=& \bib{}{x^1} + \partial_{2} \psi \bib{}{\tau} \\
  V_3 &=& \bib{}{x^2} - \partial_{1} \psi \bib{}{\tau},
\end{eqnarray}
If the function $\psi$ is independent of $\tau$ and satisfies the
3-dimensional continual Toda equation:
\begin{equation}
  {\partial_{1}}^2 \psi + {\partial_{2}}^2 \psi + 
 {\partial_{3}}^2 e^{\psi}=0,
\end{equation}
these vector fields are solutions of the half-flat equations (\ref{eq:MN1}) and
(\ref{eq:MN2}). Then, the hyperk\"ahler metric (the real heaven solution) is
given by~\cite{B-F}
\begin{equation}
ds^2=(\partial_3 \psi )^{-1} (d\tau +\beta)^2 
+(\partial_3 \psi )\gamma_{ij}dx^i dx^j, 
\end{equation}
where
\begin{equation}
\beta = -\partial_2 \psi dx^1 +\partial_1 \psi dx^2,
\end{equation}
and $\gamma_{ij}$ is the diagonal metric
$\gamma_{11}=\gamma_{22}=e^{\psi},\gamma_{33}=1$. 

In this case we find asolution of (\ref{eq:liedelT}) and (\ref{eq:VVT}):
\begin{equation}
T=c_1 (\partial_1 \psi)\bib{}{\tau}+c_2 (\partial_2 \psi) \bib{}{\tau} \quad
\mbox{for constants}\  c_i \ (i=1,2).  
\end{equation}
The corresponding ASD Maxwell solution is given by
\begin{equation}
b=c_1 da^{(1)} +c_2 da^{(2)} ,
\end{equation}
where
\begin{eqnarray}
a^{(1)} &=& \partial_1 \psi (\partial_3 \psi )^{-1}(d\tau + \beta)
+\partial_3 e^{\psi} dx^2 - \partial_2 \psi dx^3, \\
a^{(2)} &=& \partial_2 \psi (\partial_3 \psi )^{-1}(d\tau + \beta)
-\partial_3 e^{\psi} dx^1 + \partial_1 \psi dx^3. 
\end{eqnarray}
\section*{Acknowledgments}
We want to thank Y. Hashimoto for many useful discussions.


\begin{thebibliography}{99}

\bibitem{A-J-S}A. Ashtekar, T. Jacobson and L. Smolin, 
A new characterization of half-flat solutions to Einstein's equation, 
Commun. Math. Phys. 115 (1988) 631-648.

\bibitem{M-N}L.J. Mason and E.T. Newman, 
A connection between the Einstein and Yang-Mills equations, 
Commun. Math. Phys. 121 (1989) 659-668.

\bibitem{Hashi}Y. Hashimoto, Y. Yasui, S. Miyagi and T. Ootsuka,
Applications of the Ashtekar gravity to four-dimensional
hyperk\"ahler geometry and Yang-Mills instantons,
J. Math. Phys. 38 (1997) 5833-5839.

\bibitem{K-N-G}S.V. Ketov, H. Nishino and S.J. Gates Jr.,
Self-dual supersymmetry and supergravity in Atiyah-Ward space-time,
Nucl. Phys. B 393 (1993) 149-210.

\bibitem{G-N-K}S.J. Gates Jr., H. Nishino and S.V. Ketov,
Extended supersymmetry and self-duality in 2+2 dimensions,
Phys. Lett. B 297 (1992) 99-104.

\bibitem{C-D-J-M}R. Capovilla, J. Dell, T. Jacobson and L. Mason,
Self-dual 2-forms and gravity,
Class. Quantum Grav.8 (1991) 41-57.

\bibitem{Don}S.K. Donaldson, 
Complex cobordism, Ashtekar's equations and diffeomorphisms, in:
{\em Symplectic Geometry}, ed. D. Salamon,
London Math. Soc. (1992) 45-55. 

\bibitem{Robin}D.C. Robinson,
A $GL(2,\mathbb{C})$ formulation of Einstein-Maxwell theory,
Class. Quantum Grav. 11 (1994) L157-L161.

\bibitem{Kun-San}H. Kunitomo and T. Sano,
The Ashtekar Formulation for Canonical N=2 Supergravity,
Prog. Theor. Phys. Supplement 114 (1993) 31-39.

\bibitem{Ezawa}K. Ezawa,
Ashtekar's Formulation for N=1,2 Supergravities as "Constrained" BF Theories,
Prog. Theor. Phys. 95 (1996) 863-882.

\bibitem{O-V1}H. Ooguri anf C. Vafa,
Self-Dual and N=2 String Magic,
Mod. Phys. Lett. (1990) 1389-1398.

\bibitem{O-V2}H. Ooguri and C. Vafa,
Geometry of N=2 String,
Nucl. Phys. B361 (1991) 469-518.

\bibitem{Joy}D.D. Joyce, 
Explicit construction of self-dual 4-manifolds,
Duke Math. J. 77 No.3 (1995) 519-552.

\bibitem{G-H}G.W. Gibbons and S.W. Hawking, 
Gravitational multi-instantons,
Phys. Lett. B 78 (1978) 430-432.

\bibitem{E-H}T. Eguchi and A.J. Hanson,
Self-Dual Solutions to Euclidean Gravity,
Ann. Phys. 120 (1979) 82-106. 

\bibitem{B-F}C. Boyer and J. Finley, 
Killing vectors in self-dual, Euclidean Einstein spaces, 
J. Math. Phys. 23 (1982) 1126-1130.

\end{thebibliography}
\end{document}